\pdfoutput=1

\documentclass[11pt]{article}

\usepackage[T1]{fontenc}
\usepackage[utf8]{inputenc}
\usepackage{lmodern}
\usepackage[margin=1in]{geometry}
\usepackage{amsmath,amssymb}
\usepackage{booktabs}
\usepackage{array}
\usepackage{enumitem}
\usepackage{graphicx}
\usepackage{tikz}
\usetikzlibrary{arrows.meta,positioning,fit,backgrounds,calc}
\usepackage[hidelinks]{hyperref}
\usepackage{url}
\usepackage{microtype}

\newcommand{\yes}{Y}
\newcommand{\partly}{$\sim$}
\newcommand{\nope}{--}
\newcommand{\us}{\textmu s}

\title{Delegation Without Trust: An Empirical Gap Analysis of\\
Identity, Authorization, and Runtime Governance in\\
Multi-Agent LLM Systems}

\author{
  Panduranga Sai Varma Dantuluri\\
  VotalAI\\
  \texttt{sai@votal.ai}\\
  ORCID: 0009-0007-3658-3681
  \and
  Jyotirmoy Sundi\\
  VotalAI\\
  \texttt{sundi@votal.ai}
}

\date{August 31, 2026}

\begin{document}
\maketitle

\begin{abstract}
Autonomous LLM agents increasingly act on a user's behalf: they hold
credentials, call tools and remote services, and spawn sub-agents that act
further on their behalf. This turns a long-standing distributed-systems
question -- who is authorized to do what, on whose authority -- into an urgent
and largely unsolved problem, because the component driving each agent is a
language model that can be hijacked by adversarial input. We argue that agent
security must be evaluated under an untrusted-model assumption: a correct
system is one in which a fully prompt-injected agent still cannot exceed the
authority explicitly delegated to it. Against this standard we make three
contributions. First, we give a threat model for multi-agent delegation
centered on four concrete adversaries -- confused deputy, token theft and
replay, prompt-injection privilege escalation, and compromised sub-agents --
and derive eight security requirements a governed agent system must meet.
Second, we show the gap is real: a default agent runtime modeling common
framework practice (broad bearer credentials, authorization gated inside the
model) fails all four threats, and across four widely used frameworks --
LangGraph, CrewAI, AutoGen, and the Model Context Protocol (MCP)
authorization model -- three provide no built-in confinement and one only
partial; no existing standard, alone, covers the requirement set. Third, we
implement and adversarially evaluate an authorization broker that closes the
gap. It blocks all four threats; it resists 11 direct attacks on its own
design and accepts 0 of 200,000 forged tokens; it confines a compromised
sub-agent to its delegated task (a mean of 1.5 reachable actions versus all
8,100 under bearer delegation, across 2,000 randomized scenarios); and it
enforces at microsecond cost ($\sim$2.6\,\us{} per decision), negligible
against model inference. These principles are additionally realized in a
production system, VotalAI's LLM Shield.
\end{abstract}

\section{Introduction}

An LLM agent is no longer a chatbot; it is a process that holds credentials
and takes consequential actions -- reading mailboxes, moving money, filing
tickets, calling internal APIs -- on a human's behalf. In multi-agent
deployments a primary agent further delegates to specialized sub-agents, each
of which may hold its own credentials and call its own
tools~\cite{autogen,agentsurvey}. The security question this raises is ancient
and well studied in distributed systems: who is authorized to perform which
action, on whose authority, and how is that authority scoped, proven, rotated,
and revoked? What is new, and what existing identity infrastructure was never
designed for, is that the entity exercising the authority is a language model
whose behavior an adversary can commandeer through direct or indirect prompt
injection~\cite{greshake,injecagent}.

The prevailing engineering posture treats the model as a trusted component:
agent frameworks hand long-lived bearer credentials to tools, delegate to
sub-agents without narrowing scope, and let the model's own reasoning gate
sensitive actions. We argue this is backwards. Because the model is the most
attackable part of the system, it must be treated as untrusted, and the
correct evaluation standard is adversarial:

\paragraph{Untrusted-model property.}
A multi-agent system is well-governed only if a fully prompt-injected agent --
one whose model output is entirely attacker-controlled -- still cannot exceed
the authority explicitly delegated to it, cannot use a stolen credential off
its own workload, and leaves an attributable audit trail. Security that
depends on the model not being hijacked is not security.

We hold the entire paper to this standard. Our contributions:

\begin{enumerate}[leftmargin=*]
\item A \textbf{threat model for multi-agent delegation} (Section~\ref{sec:threat})
centered on four adversaries -- confused deputy, token theft/replay,
prompt-injection privilege escalation, and compromised sub-agents -- under the
untrusted-model assumption, and eight derived security requirements
(Section~\ref{sec:reqs}).

\item A \textbf{gap analysis} (Sections~\ref{sec:gap},~\ref{sec:systematization}):
a default agent runtime fails all four threats; across four widely used
frameworks, three provide no built-in confinement and one only partial; and no
existing standard, alone, covers the requirement set -- the primitives have not
been composed for agent-to-agent delegation.
Concurrent work~\cite{scopegate} independently reports the absence of per-call
authorization in an overlapping set of frameworks; we situate our analysis
against it in Section~\ref{sec:related}.

\item An \textbf{implemented, adversarially-evaluated broker}
(Section~\ref{sec:broker}) that composes the primitives and closes the gap: it
blocks all four threats, resists 11 direct attacks on its design and 0 of
200,000 forged tokens, confines a compromised sub-agent to its task, and
enforces at microsecond cost.
\end{enumerate}

This is a problem-and-solution paper: we demonstrate the gaps rather than
asserting them, and we build and attack the defense rather than only sketching
it. Our evaluation is built entirely on public frameworks and open standards;
Section~\ref{sec:shield} describes how the same principles are realized in
production in VotalAI's LLM Shield.

\section{Background}

\paragraph{Agentic architectures.}
Modern agent frameworks~\cite{autogen} orchestrate an LLM in a
plan--act--observe loop with access to tools. Multi-agent variants add
delegation: a primary agent decomposes a task and assigns sub-tasks to other
agents, forming a delegation chain rooted, ideally, in a human grant. The
Model Context Protocol (MCP)~\cite{mcpspec,mcpauthz} standardizes how agents
connect to tools and servers and defines an authorization profile built on
OAuth.

\paragraph{Foundational access-control principles.}
Least privilege and complete mediation~\cite{saltzer} and the capability
model's answer to the confused-deputy problem~\cite{hardy,capmyths} are the
lens we apply: authority should be represented as unforgeable, narrowly scoped,
delegable tokens, and every access should be mediated by a reference monitor
the subject cannot bypass.

\paragraph{Identity and delegation standards.}
OAuth 2.1~\cite{oauth21} and OpenID Connect~\cite{oidc} anchor human identity;
OAuth Token Exchange (RFC 8693)~\cite{rfc8693} expresses on-behalf-of
delegation, minting a downstream token that records the delegation chain.
SPIFFE/SPIRE~\cite{spiffe,spire} provides cryptographic workload identity (an
SVID) with automatic rotation. Sender-constraining binds a token to its holder
so a stolen token is useless elsewhere, via mutual-TLS (RFC
8705)~\cite{rfc8705} or DPoP (RFC 9449)~\cite{rfc9449}.
Macaroons~\cite{macaroons} and Biscuit~\cite{biscuit} are capability tokens
that support attenuation: any holder can add caveats that narrow scope, but
none can widen it -- a natural fit for delegation chains. Zero-trust
architecture~\cite{nist800207} frames the overall posture. Each primitive is
mature; the open question is whether they compose to govern untrusted agents,
which Sections~\ref{sec:systematization} and~\ref{sec:broker} address.

\section{Threat Model}
\label{sec:threat}

\paragraph{Principals.}
A human user (the root of authority, authenticated via an OIDC identity
provider); a primary agent; one or more sub-agents; tools / MCP servers; and
the authorization infrastructure (identity provider, token service, and policy
enforcement points, PEPs).

\paragraph{Trust assumptions.}
The authorization infrastructure and the PEPs are trusted and correctly
implemented, and workload identities are authenticated (e.g., via mutual-TLS /
SPIFFE attestation). \emph{Every agent's model is untrusted}: an adversary may
fully control any agent's model output via direct or indirect prompt
injection~\cite{greshake}. A network adversary may observe and replay traffic
and may exfiltrate a credential from a compromised agent or tool. We do not
assume the adversary breaks cryptography or compromises the infrastructure
itself. Section~\ref{sec:broker} shows this authenticated-identity assumption
is load-bearing: sender-constraining is only as strong as workload-identity
attestation.

\paragraph{Security goal.}
Under these assumptions: (G1) a hijacked agent cannot exercise authority
beyond what was explicitly delegated to it; (G2) each delegation hop can only
narrow authority; (G3) a stolen credential cannot be used off the workload it
was issued to; (G4) a compromised agent's authority can be promptly revoked;
and (G5) every privileged action is attributable to a delegation chain.

\paragraph{Adversaries.}

\begin{description}[leftmargin=1.5em,style=nextline]
\item[T1 -- Confused deputy.]
An agent holding legitimate authority is manipulated, via injected content,
into exercising that authority toward the attacker's goal~\cite{hardy}. The
agent is authorized; the use is not intended.

\item[T2 -- Token theft / replay.]
A bearer credential is exfiltrated from an agent or tool and replayed by the
attacker from a different context.

\item[T3 -- Prompt-injection privilege escalation.]
Injected content causes an agent to acquire or exercise permissions beyond
those its task requires (e.g., requesting a broader scope, invoking an
out-of-scope tool).

\item[T4 -- Compromised sub-agent / over-broad scope.]
A sub-agent (or one of its tools) is compromised; because it holds over-broad
or non-attenuated delegated authority, its blast radius exceeds its task.
\end{description}

\section{Security Requirements}
\label{sec:reqs}

From the goals we derive the requirements a governed multi-agent system must
satisfy. Each names the threats it addresses.

\begin{description}[leftmargin=1.5em,style=nextline]
\item[R1 Delegation-chain integrity / on-behalf-of provenance.]
Every hop's authority is cryptographically traceable to a human grant.
[T1,\,T4]

\item[R2 Least-privilege attenuation.]
Each delegation can only narrow scope, never widen it. [T1,\,T3,\,T4]

\item[R3 Sender-constrained credentials.]
Tokens are bound to a workload identity, so exfiltration yields nothing
usable. [T2]

\item[R4 Short-lived issuance.]
Credentials are minted with a minimal lifetime, shrinking the misuse window.
[T2,\,T3]

\item[R5 Key/secret rotation.]
Workload keys and secrets rotate automatically, bounding long-term compromise.
[T2]

\item[R6 Revocation propagation.]
Authority can be revoked and the revocation reaches PEPs promptly. [T1,\,T4]

\item[R7 Complete, tamper-evident audit.]
Every privileged action is attributable to a delegation chain.
[detection/forensics, all]

\item[R8 Model-independent enforcement.]
Authorization decisions are made by infrastructure PEPs, never by the
untrusted model. [all -- the crux]
\end{description}

R8 is the load-bearing requirement: it is what makes the untrusted-model
property achievable at all. If the model gates access, a hijacked model grants
access.

\section{Empirical Gap Analysis}
\label{sec:gap}

\subsection{A default runtime fails all four threats}

We first establish the baseline. We implement a default runtime modeling
common agent-framework practice: a single broad bearer credential is issued to
the primary agent and passed unchanged to sub-agents and tools, and whether an
action is permitted is decided by the agent's own reasoning (model-gated), not
by an external policy point. Running the four adversaries against this
runtime, a hijacked agent succeeds at all four: the confused-deputy misuse, the
replayed token, the escalated privilege, and the over-broad sub-agent all go
through, because the runtime provides none of R1--R8. Our implementation
reproduces this.

\subsection{Four frameworks}
\label{sec:frameworks}

We evaluate four widely used frameworks and record, per threat, whether each
framework's defaults prevent it (\emph{vulnerable} = defaults do not prevent
it; \emph{partial}; \emph{mitigated}), and which of R1--R8 the default
identity/authorization story provides. Our methods differ by framework and we
state them explicitly. For LangGraph we \emph{execute} the scenario: we drive a
real compiled StateGraph with a scripted agent node (isolating the
authorization layer from the model, since we test authority confinement rather
than model behavior) and observe the tool node. For CrewAI and AutoGen we
perform \emph{capability inspection} of the frameworks' tool-execution base
classes. For MCP we perform \emph{specification analysis} of the authorization
profile. This is capability analysis on the same footing as the standards
table (Section~\ref{sec:systematization}); all four rows are grounded in an
execution adapter for LangGraph and an inspection of the CrewAI/AutoGen code
rather than asserted.

\begin{table}[t]
\centering
\caption{Threat outcomes by framework (default configuration).
\yes{} = vulnerable, \partly{} = partial/only with hardening,
\nope{} = mitigated by default. LangGraph executed; CrewAI/AutoGen inspected;
MCP from spec.}
\label{tab:threats}
\begin{tabular}{lcccc}
\toprule
Framework & T1 deputy & T2 replay & T3 escal. & T4 sub-agent\\
\midrule
LangGraph 1.2.10        & \yes & \yes    & \yes & \yes\\
CrewAI 1.15.13          & \yes & \yes    & \yes & \yes\\
AutoGen 0.7.5           & \yes & \yes    & \yes & \yes\\
MCP (authz 2026-07-28)  & \yes & \partly & \yes & \yes\\
\bottomrule
\end{tabular}
\end{table}

\begin{table}[t]
\centering
\caption{Default requirement coverage per framework. \yes{} = provides by
default; \partly{} = partial; \nope{} = absent. Three frameworks provide none;
MCP's OAuth model gives partial coverage.}
\label{tab:coverage}
\begin{tabular}{lcccccccc}
\toprule
Framework & R1 & R2 & R3 & R4 & R5 & R6 & R7 & R8\\
\midrule
LangGraph 1.2.10       & \nope   & \nope & \nope   & \nope & \nope   & \nope   & \nope & \nope\\
CrewAI 1.15.13         & \nope   & \nope & \nope   & \nope & \nope   & \nope   & \nope & \nope\\
AutoGen 0.7.5          & \nope   & \nope & \nope   & \nope & \nope   & \nope   & \nope & \nope\\
MCP (authz 2026-07-28) & \partly & \nope & \partly & \yes  & \partly & \partly & \nope & \partly\\
\bottomrule
\end{tabular}
\end{table}

The result: three of the four -- LangGraph, CrewAI, and AutoGen -- provide no
built-in authorization or delegation confinement (none of R1--R8). In their
default configuration a hijacked agent's out-of-scope tool calls execute, and
confinement is left to the developer; none claims to be an authorization
system, which is precisely the point -- the governance layer is simply absent.
Concurrent and independent work~\cite{scopegate} reaches a convergent
conclusion for a partially overlapping framework set (LangChain/LangGraph,
LlamaIndex, and the Stripe Agent Toolkit), reporting that none re-authorizes
model-emitted calls against concrete argument values by default. We view the
agreement as corroboration of the finding; Section~\ref{sec:related} details
how the two analyses differ in threat model, framework and standards coverage,
and evaluation.
MCP is the exception: its authorization specification (rev.\ 2026-07-28) is an
OAuth 2.1 resource-server model, giving partial coverage. Its mandatory
Resource Indicators (RFC 8707) audience-restrict each token to a specific
server, preventing cross-server token reuse (partial coverage for T2 and R3);
its resource-server model validates tokens outside the client's model (R8); and
OAuth provides short-lived tokens (R4) and revocation (R6). It provides,
however, no per-hop attenuation (R2) and no cross-agent delegation provenance
(R1) for agent-to-agent chains. Tables~\ref{tab:threats}
and~\ref{tab:coverage} report the outcomes, with the MCP cells derived from the
current MCP authorization specification.

\section{Systematization: Standards vs.\ Requirements}
\label{sec:systematization}

Table~\ref{tab:standards} maps each existing standard/primitive to the
requirements it can satisfy. Like the per-framework coverage, this is an
analytical result: it reflects our reading of each standard's specification
rather than measured data, and we invite scrutiny of individual cells.

\begin{table}[t]
\centering
\caption{Which requirement each standard can satisfy. \yes{} = directly
provides; \partly{} = partial / with profiling; \nope{} = out of scope.
Derived from the specifications; the mapping reflects the authors' reading of
each standard.}
\label{tab:standards}
\begin{tabular}{lcccccccc}
\toprule
Standard / primitive & R1 & R2 & R3 & R4 & R5 & R6 & R7 & R8\\
\midrule
OIDC~\cite{oidc}                     & \partly & \nope   & \nope   & \yes    & \nope & \partly & \partly & \nope\\
OAuth 2.1~\cite{oauth21}             & \partly & \partly & \nope   & \yes    & \nope & \partly & \partly & \yes\\
Token Exchange 8693~\cite{rfc8693}   & \yes    & \partly & \nope   & \yes    & \nope & \nope   & \partly & \nope\\
SPIFFE/SPIRE~\cite{spiffe,spire}     & \partly & \nope   & \yes    & \yes    & \yes  & \partly & \nope   & \nope\\
mTLS 8705~\cite{rfc8705}             & \nope   & \nope   & \yes    & \nope   & \nope & \nope   & \nope   & \nope\\
DPoP 9449~\cite{rfc9449}             & \nope   & \nope   & \yes    & \nope   & \nope & \nope   & \nope   & \nope\\
Macaroons~\cite{macaroons}           & \yes    & \yes    & \partly & \partly & \nope & \partly & \nope   & \yes\\
Biscuit~\cite{biscuit}               & \yes    & \yes    & \partly & \partly & \nope & \partly & \partly & \yes\\
MCP authz~\cite{mcpauthz}            & \partly & \nope   & \partly & \yes    & \nope & \nope   & \partly & \partly\\
\bottomrule
\end{tabular}
\end{table}

Two observations follow. First, no single standard covers the requirement set:
delegation provenance (R1) and attenuation (R2) live in token-exchange and
capability tokens; sender-constraining (R3) and rotation (R5) live in
SPIFFE/mTLS/DPoP; model-independent enforcement (R8) is a deployment property
none of them mandates. Second, and more importantly, the primitives have not
been composed for the agent-to-agent case: there is today no standard profile
that mints, at each delegation hop, a capability token that is simultaneously
attenuated (R2), bound to the sub-agent's workload identity (R3), short-lived
(R4), and enforced at a PEP outside the model (R8). Closing that composition is
the subject of Section~\ref{sec:broker}.

\section{The Broker: Design, Implementation, and Evaluation}
\label{sec:broker}

We compose the primitives into an authorization broker / PEP that sits between
agents and tools, and we implement and adversarially evaluate it. The broker
realizes the untrusted-model property by construction: a fully hijacked agent
can exercise only its attenuated, workload-bound grant, and a stolen token is
inert off its identity.

\subsection{Design}

\begin{figure}[t]
\centering
\begin{tikzpicture}[
  font=\small,
  box/.style={draw,rounded corners=2pt,align=center,inner sep=5pt,minimum height=1cm},
  agent/.style={box,fill=black!3},
  flow/.style={-{Latex[length=2mm]},thick},
  lbl/.style={font=\scriptsize,inner sep=1.5pt}
]

\node[box] (human) at (0,4.3) {Human\\\scriptsize OIDC identity (root)};

\node[box,minimum width=4.6cm] (broker) at (0,2.1)
  {\textbf{Broker / PEP}\\\scriptsize issuer + model-independent\\[-1pt]
   \scriptsize enforcement $\cdot$ audit $\cdot$ revocation};

\node[box] (tool) at (7.4,2.1) {Tool / MCP server};

\node[agent] (primary) at (0,-0.6) {Primary agent\\\scriptsize $\mathrm{SVID}_1$, $\mathrm{cap}_1$};

\node[agent] (sub) at (0,-2.7) {Sub-agent\\\scriptsize $\mathrm{SVID}_2$, $\mathrm{cap}_2 \subseteq \mathrm{cap}_1$};

\begin{scope}[on background layer]
  \node[draw,dashed,rounded corners,fill=black!2,
        fit=(primary)(sub),inner sep=10pt] (untrusted) {};
\end{scope}
\node[lbl,anchor=north west] at ([xshift=2pt,yshift=-3pt]untrusted.south west)
  {untrusted: model-controlled};

\node[lbl,anchor=south west] at ([xshift=6pt,yshift=3pt]broker.north east)
  {trusted: no model};

\draw[flow] (human) -- node[lbl,left] {1.\ authenticate} (broker);

\draw[flow] (broker.west) -- ++(-1.5,0) |- (primary.west);
\node[lbl,left] at (-3.85,0.75) {2.\ mint $\mathrm{cap}_1$};

\draw[flow] (primary) -- node[lbl,right] {3.\ delegate} (sub);

\draw[flow] (sub.east) -- (4.0,-2.7) -- (4.0,1.55) -- (broker.south east);
\node[lbl,right] at (4.05,-1.3) {4.\ tool call ($\mathrm{cap}_2$, $\mathrm{SVID}_2$)};

\draw[flow] (broker.east) -- node[lbl,above] {5.\ verify + enforce}
                             node[lbl,below] {allow / deny} (tool.west);

\end{tikzpicture}
\caption{Data flow through the broker. A human's OIDC identity roots a
capability token; each delegation hop is an issuer-mediated token exchange
that mints an attenuated ($\mathrm{cap}_2 \subseteq \mathrm{cap}_1$),
SVID-bound, short-lived token. Every tool call is routed through the broker,
which verifies the token and enforces its caveats outside the model (R8)
before allowing the call. Agents run in an untrusted zone; a fully hijacked
agent can exercise only its bound, attenuated grant.}
\label{fig:dataflow}
\end{figure}
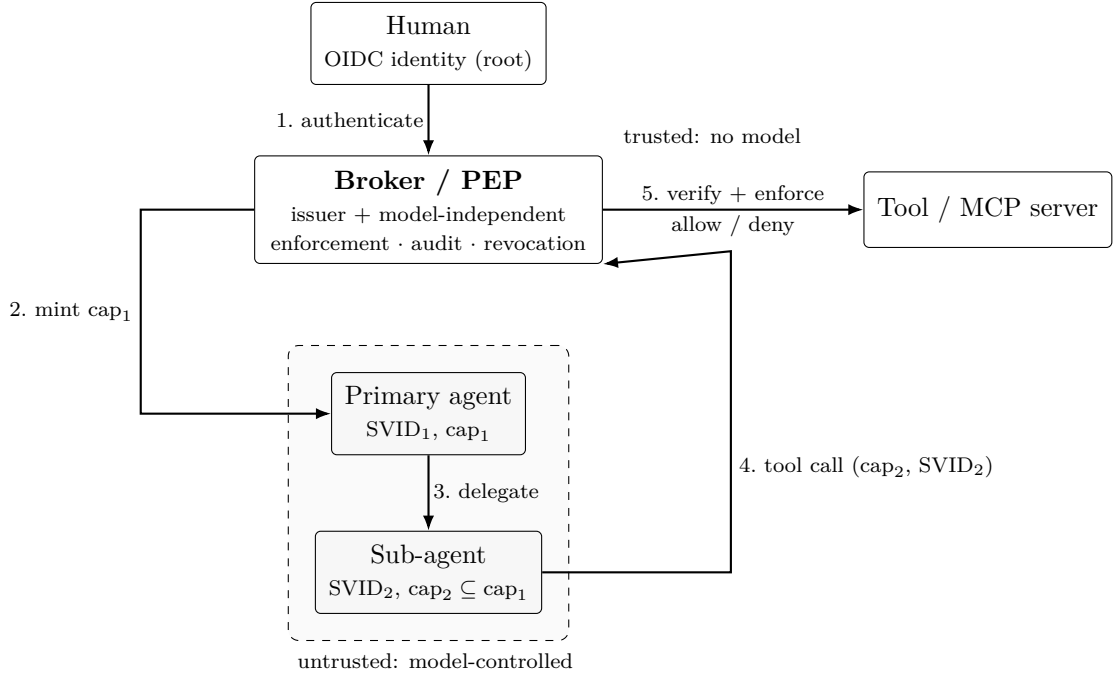

(1) Each agent authenticates with a SPIFFE-style workload identity (SVID)
(R3, R5). (2) The human grant is a capability token rooted at the OIDC
identity (R1). (3) At each delegation hop, the broker performs an
issuer-mediated token exchange that mints a new token, attenuated to the
sub-task (R2), bound to the receiving agent's SVID (R3), and short-lived (R4);
only the workload a token is bound to may delegate it onward. (4) Every tool
call is mediated by the broker, which verifies the token and enforces its
caveats independently of the model (R8) and appends the delegation chain to a
tamper-evident log (R7). (5) Revocation invalidates a token and propagates
down its chain (R6). Figure~\ref{fig:dataflow} shows the resulting data flow.

\subsection{Implementation}

We implement the broker in Python ($\sim$160 lines, standard library only).
Tokens are HMAC-signed in the style of macaroons and OAuth Token Exchange:
forgery, caveat removal, and subject re-binding require the broker secret and
therefore fail verification; caveats are append-only, so authority can only
narrow. Enforcement is a signature check, a revocation check, a
sender-constraint check, and caveat evaluation -- no model call. The default
runtime of Section~\ref{sec:gap} and the broker share a common scenario
harness.

\subsection{Evaluation}

We evaluate along four axes; all numbers are from our reference implementation.

\paragraph{Defense effectiveness.}
Against the four adversaries, the default runtime fails all four and the
broker blocks all four (Table~\ref{tab:attack}).

\begin{table}[t]
\centering
\caption{Attack outcome by runtime. The broker blocks every threat the default
runtime admits.}
\label{tab:attack}
\begin{tabular}{lcc}
\toprule
Threat & Default runtime & Broker\\
\midrule
T1 confused deputy       & succeeds & blocked\\
T2 token theft / replay  & succeeds & blocked\\
T3 privilege escalation  & succeeds & blocked\\
T4 compromised sub-agent & succeeds & blocked\\
\bottomrule
\end{tabular}
\end{table}

\paragraph{Adversarial robustness.}
Beyond the four scenarios, we run a suite that actively attacks the broker's
own design: forging a token from scratch, forging with a guessed secret,
stripping caveats, widening a caveat, re-binding a stolen token, replaying
as-is, a malformed-caveat fail-closed check, widening scope via exchange,
delegating a token one does not hold, using a child of a revoked token, and
using an expired token. The broker resists all 11. Fuzzing with random and
mutated tokens, it accepts 0 of 200,000. One boundary test succeeds by design:
if an attacker can present the victim's SVID, a stolen token works --
confirming that sender-constraining reduces to the strength of
workload-identity attestation (mTLS/SPIFFE), the assumption stated in
Section~\ref{sec:threat}. We report this as a stated assumption, not a flaw.

\paragraph{Blast radius.}
We measure how much a compromised sub-agent can reach, over 2,000 randomized
delegation scenarios in a synthetic environment of 8,100 tool/resource
actions. Under bearer delegation the compromised sub-agent reaches all 8,100
in every scenario; under the broker it reaches a mean of 1.5 actions --
exactly its delegated sub-task. The reduction is not the headline; the
invariant is: defended reach is bounded by the task and independent of
environment size, whereas undefended reach equals the environment and grows
without bound as a deployment scales.

\paragraph{Overhead.}
Enforcement costs $\sim$2.6\,\us{} per authorization ($\sim$3.9$\times$10\textsuperscript{5}
decisions/s) and a token exchange $\sim$5.4\,\us{}, measured over
2$\times$10\textsuperscript{5} calls each on a laptop. Against model inference
measured in hundreds of milliseconds to seconds, the governance layer is
effectively free.

\section{Real-World Implementation: VotalAI LLM Shield}
\label{sec:shield}

The broker of Section~\ref{sec:broker} is a minimal reference implementation
intended to isolate the requirements with the least machinery. Its principles
are realized in production in VotalAI's LLM Shield, a runtime governance layer
for agentic LLM deployments. Where the reference implementation uses
SPIFFE-style workload identity and macaroon-style attenuation to illustrate
R1--R8, Shield satisfies the same requirements through a production identity
and policy stack. The empirical results in this paper are from our reference
implementation; the architecture below is drawn from publicly presented
material, and production figures are labeled as such. VotalAI LLM Shield is a
commercial product; the reference implementation evaluated in this paper is a
separate, minimal demonstrator built on public primitives and is not released.

\paragraph{Architecture.}
Client, RAG, MCP, and agentic applications connect through a single AI Gateway
that provides authentication, authorization, routing, rate limiting, and
observability, and embeds an MCP gateway (built on components such as Kong,
LiteLLM, and Portkey). Agent identity is issued and validated by the VotalAI
Agent IDP, which federates with enterprise identity providers (Okta, Auth0,
Microsoft Entra, Google) and offers OIDC, SSO, MFA, attribute-based access
control (ABAC), and a policy enforcement point; agents authenticate to tools
and MCP servers using IDP-issued tokens. Requests traverse guardrail model
services containing pre-call and post-call content inspection and --- central
to this paper --- an Agent \& Tool Access Control stage performing role-based
tool authorization and per-agent permission enforcement. A Redis tier caches
tenant policy for fast lookups, and an admin/tenant portal manages policy. The
content-inspection layer combines VotalAI's own guardrail model with NVIDIA's
separate Nemotron 3.5 content-safety model (two distinct models); this layer is
complementary to the authorization layer that is the focus of this paper,
consistent with our position (Section~\ref{sec:discussion}) that content
filtering cannot replace authority scoping.

\paragraph{Mapping to the requirements.}
Shield realizes the requirements through this stack rather than through the
reference primitives. On-behalf-of provenance (R1) and short-lived credential
issuance (R4) come from the Agent IDP's OIDC token service; least-privilege
attenuation (R2) is enforced by ABAC and the Agent \& Tool Access Control
stage's role-based tool authorization; model-independent enforcement (R8) ---
the load-bearing requirement --- is provided by the gateway and access-control
policy enforcement points, which sit outside the model on the request path;
and complete audit (R7) is provided by the platform's audit logs, metrics, and
policy-decision records. Sender-constrained credentials (R3), key rotation
(R5), and revocation (R6) are inherited from the federated enterprise identity
providers. The design follows zero-trust and least-privilege principles with
end-to-end encryption throughout.

\paragraph{Deployment and scale.}
Shield ships as a container image and scales through container replicas for
multi-tenant deployment; the policy tier runs Redis for tenant policy mapping,
and the guardrail model services run vLLM inference of VotalAI's 8B guardrail
model (fine-tuned on VotalAI's adversarial dataset) and NVIDIA's separate
Nemotron 3.5 content-safety model on NVIDIA H100/B200 GPUs (80\,GB VRAM). As
production figures for the content-inspection layer (distinct from, and far
larger than, the reference-harness authorization overhead of
$\sim$2.6\,\us{}), pre-call inspection adds on the order of 2\,ms and a full
guardrail-model check on the order of 190\,ms per request. This separation is
itself informative: authorization enforcement is comparatively negligible
(Section~\ref{sec:broker}), whereas the dominant runtime cost is the
content-safety model inference --- a complementary layer, not a substitute for
authority scoping.

\paragraph{Continuous red-teaming.}
Shield includes an attack zone / red-teaming portal that continuously probes
protected endpoints, synthesizing adversarial prompts on the fly in
single-turn, multi-turn, and combined modes from a catalog of over 100
manipulation strategies --- an operational analogue of the adversarial
evaluation in Section~\ref{sec:broker}.

\section{Discussion}
\label{sec:discussion}

\paragraph{Why model-side guardrails are not enough.}
Input/output content filters reduce the rate of successful injection but
cannot provide R1--R8; they are probabilistic and sit inside the trust
boundary we assume broken. Governance must be enforced by authority scoping
outside the model, which content filtering complements but cannot replace.

\paragraph{Deployment cost.}
The broker adds a mediation point on the tool-call path and a per-hop token
exchange. As Section~\ref{sec:broker} shows, enforcement is microseconds and
token minting a few microseconds -- orders of magnitude below the model call it
accompanies -- so the dominant cost is operational (running a broker and
identity plane), not latency.

\section{Threats to Validity}

\emph{Construct}: the blast-radius magnitude depends on the ratio of task
scope to environment size; we therefore report absolute reachable counts and
the size-independence invariant rather than a single percentage.
\emph{Internal}: our broker enforces an abstract action model (tool/resource
tuples); a production PEP must map real tool invocations to this model
faithfully. \emph{External}: the per-framework results are established for
specific, pinned framework versions (Tables~\ref{tab:threats},
\ref{tab:coverage}), by execution for LangGraph and by capability inspection
of code/spec for CrewAI, AutoGen, and MCP; frameworks evolve, and the MCP
cells in particular should be re-checked against the current authorization
spec. \emph{Scope}: we implement and adversarially evaluate the broker but do
not yet integrate it into the evaluated frameworks; that integration, and an
end-to-end evaluation with a live model in the loop, are future work.

\section{Ethics and Responsible Disclosure}

All demonstrations target public agent frameworks in a test harness under our
control; none attack third-party or production systems, and no user data is
involved. The demonstrated weaknesses are the absence of authorization
primitives in developer frameworks rather than exploitable defects; we
nonetheless plan to notify the relevant framework maintainers of the coverage
findings through their published channels. The reference implementation
described here is not publicly released; this paper contains no proprietary
LLM Shield internals or client data, and the production description in
Section~\ref{sec:shield} is limited to details cleared for publication.

\section{Related Work}
\label{sec:related}

\paragraph{Attacks and measurement.}
Prompt injection against LLM-integrated applications and agents is documented
by Greshake et al.~\cite{greshake} and, for tool-integrated agents, by
InjecAgent~\cite{injecagent}, and evaluated at scale by
AgentDojo~\cite{agentdojo} and the sandboxed-agent risk study of Ruan et
al.~\cite{toolemu}. Abdelnabi and Bagdasarian~\cite{alwaysfall} push this
further, arguing that agents may always remain susceptible to injection, so
that any defense resting on the model's own compliance is resting on a moving
floor. We take that conclusion as our starting premise rather than a result to
be re-established: our untrusted-model assumption grants the attacker full
control of model output and asks the orthogonal question of whether the
identity and authorization layer confines the damage. On the deployment side,
Hou et al.~\cite{mcplandscape} give a systematic threat taxonomy for MCP, and
Zhou et al.~\cite{mcpauthmeasure} measure authentication in the wild across
7,973 live remote MCP servers, finding that over 40\% expose tools with no
authentication at all and that every one of the 119 OAuth-enabled servers they
tested carried at least one authentication flaw. Their measurement is the
field-scale complement to our specification-level analysis of the MCP
authorization profile (Section~\ref{sec:frameworks}): they establish that
deployed MCP authorization is frequently absent or misconfigured, while we ask
what the specification would fail to cover even when implemented correctly.

\paragraph{Defenses at the content and data-flow layer.}
A line of work removes or neutralizes the injection vector before the model
acts on it. Untrusted Content Masking~\cite{ucm} masks untrusted regions of a
web page so the agent never processes injected text, and explicitly notes
residual data-flow attacks -- where a compromised model is steered by
trusted-looking data rather than injected instructions -- as outside the scope
of control-flow defenses. CaMeL~\cite{camel} extracts control and data flow
from the trusted user query and confines the agent to that plan with
capability-tagged data, and FIDES~\cite{fides} carries confidentiality and
integrity labels through the agent's execution and enforces information-flow
policies deterministically outside the model. These bound what an agent may
\emph{know} and which data may influence which action. We bound what an agent
may \emph{do} on whose authority. The two axes are complementary and neither
subsumes the other: content masking removes the injection vector,
information-flow control constrains propagation, and attenuated,
model-independent authorization bounds the blast radius of an agent that has
already been subverted -- precisely the residual case~\cite{ucm} names.

\paragraph{Enforcement outside the model.}
Closest to our position is a growing body of work that moves the authorization
decision out of the model and into infrastructure. Progent~\cite{progent}
gives each agent a programmable privilege policy over tool calls and enforces
it deterministically, with the property that the permitted action space can
only narrow without explicit user approval. FORGE~\cite{forge} treats policy
enforcement as a cross-cutting concern, specifying Datalog policies
independently of the agent's reasoning and consulting a reference monitor at
every policy-relevant decision under a formal assume/guarantee contract.
PAuth~\cite{pauth} binds authority to the concrete task the user asked for,
deriving symbolic slices from the natural-language request so that a server
can verify an operand was legitimately produced rather than injected.
AC4A~\cite{ac4a} observes that agent access to APIs is today all-or-nothing
and proposes finer-grained scoping. We share the central commitment of all
four -- our R8 -- and differ on layer and on scope. These systems are policy
engines for a single agent's tool calls: they answer \emph{is this call
permitted}. Our concern is the credential and identity layer underneath that
question, across delegation hops: whether the authority a sub-agent presents is
cryptographically traceable to a human grant (R1), can only narrow at each hop
(R2), is bound to the presenting workload so a stolen token is inert (R3), and
can be rotated and revoked (R5, R6). A policy engine that decides correctly on
a forged, replayed, or over-broad delegated credential still authorizes the
wrong principal. The two compose naturally: a broker of the kind we build
supplies the attenuated, sender-constrained authority that a policy engine such
as FORGE or Progent then evaluates.

\paragraph{Concurrent framework audits.}
Concurrent and independent work by Mellafe Zuvic~\cite{scopegate} audits
whether LangChain/LangGraph, LlamaIndex, and the Stripe Agent Toolkit
re-authorize each model-emitted call with concrete argument values before
execution, finds that all three provide capability gating but none provides a
deterministic fail-closed per-call authorization gate by default, and proposes
ScopeGate, a five-stage policy decision/enforcement point. That work and ours
reach the same conclusion about LangChain/LangGraph defaults from different
starting points and with different methods, which we regard as mutual
corroboration of the central empirical claim rather than a duplication of it.
The analyses differ in scope in three ways. First,~\cite{scopegate} evaluates
the single-agent tool-call path; our threat model (Section~\ref{sec:threat}) is
multi-agent delegation, and three of our four adversaries -- token
theft/replay, privilege escalation across a hop, and compromised sub-agent --
have no analogue in a single-hop gate. Second,~\cite{scopegate} audits three
framework code bases; we cover LangGraph, CrewAI, AutoGen, and the MCP
authorization specification, and pair the framework analysis with a
systematization of nine identity and delegation standards against a
requirements framework (Section~\ref{sec:systematization}), which locates the
gap in the standards landscape rather than only in framework defaults. Third,
ScopeGate's evaluation measures denial of unauthorized calls; ours additionally
measures properties specific to delegated authority -- resistance to token
forgery across 200,000 attempts, and the blast radius of a compromised
sub-agent across randomized delegation scenarios -- which are the quantities
that distinguish attenuated delegation from a per-call allowlist.

\paragraph{Governance frameworks and primitives.}
Community efforts enumerate and govern agentic risk: the OWASP LLM Top
10~\cite{owaspllm}, and -- for autonomous agents specifically -- the OWASP
Agentic Security Initiative's Top 10 for Agentic
Applications~\cite{owaspagentic}, its Agentic AI Threats and Mitigations
taxonomy~\cite{owaspthreats}, and its State of Agentic AI Security and
Governance report~\cite{owaspstate}. These catalogue risks and frame
governance; we complement them with an empirical, standards-grounded gap
analysis, a requirements framework, and an implemented,
adversarially-evaluated defense. The building blocks -- capability
security~\cite{hardy,capmyths,macaroons,biscuit},
delegation~\cite{rfc8693}, workload identity~\cite{spiffe}, and
sender-constraining~\cite{rfc8705,rfc9449} -- are mature; our contribution is
to show they are neither composed nor shipped for untrusted multi-agent
delegation, and to compose them in a broker that meets the requirements.

\section{Conclusion}

Agent frameworks today trust the least trustworthy component in the system --
the model -- to gate authority. Evaluated under an untrusted-model standard, a
default runtime fails all four core threats of multi-agent delegation; three of
four popular frameworks ship no confinement and the fourth only partial; and no
existing standard alone closes the gap. The primitives to do better exist; they
had simply not been composed for agent-to-agent, on-behalf-of authority. We
composed them in a broker, attacked it, and showed it confines a hijacked agent
to its delegated task at microsecond cost. We hope ``does this hold when the
model is hijacked?'' becomes the default question asked of every agent
deployment.

\end{document}